\documentclass[sigconf]{acmart}
\usepackage{balance}
\usepackage{booktabs} % For formal tables
\usepackage[]{algorithm2e}
\usepackage{tabularx}
\usepackage{multirow}
\usepackage{subfigure}
\usepackage{CJKutf8}
\usepackage[utf8]{inputenc}
\usepackage[T1]{fontenc}

\def\BibTeX{{\rm B\kern-.05em{\sc i\kern-.025em b}\kern-.08emT\kern-.1667em\lower.7ex\hbox{E}\kern-.125emX}}
    
\copyrightyear{2019}
\acmYear{2019}
\setcopyright{acmcopyright}
\acmConference[KDD '19]{The 25th ACM SIGKDD Conference on Knowledge Discovery and Data Mining}{August 4--8, 2019}{Anchorage, AK, USA}
\acmBooktitle{The 25th ACM SIGKDD Conference on Knowledge Discovery and Data Mining (KDD '19), August 4--8, 2019, Anchorage, AK, USA}
\acmPrice{15.00}
\acmDOI{10.1145/3292500.3330727}
\acmISBN{978-1-4503-6201-6/19/08}

\begin{document}

% The "title" command has an optional parameter, allowing the author to define a "short title" to be used in page headers.
\title{A User-Centered Concept Mining System for Query and Document Understanding at Tencent}

\author{Bang Liu$^{1*}$, Weidong Guo$^{2*}$, Di Niu$^1$, Chaoyue Wang$^2$}
\author{Shunnan Xu$^2$, Jinghong Lin$^2$, Kunfeng Lai$^2$, Yu Xu$^2$}
\thanks{$^*$These authors contributed equally to this work.}
\affiliation{$^1$University of Alberta, Edmonton, AB, Canada}
\affiliation{$^2$Platform and Content Group, Tencent, Shenzhen, China}
%\email{haolan@ualberta.ca}
%\email{bang3@ualberta.ca}

%
% The abstract is a short summary of the work to be presented in the article.
\begin{abstract}
Concepts embody the knowledge of the world and facilitate the cognitive processes of human beings.
Mining concepts from web documents and constructing the corresponding taxonomy are core research problems in text understanding and support many downstream tasks such as query analysis, knowledge base construction, recommendation, and search. However, we argue that most prior studies extract formal and overly general concepts from Wikipedia or static web pages, which are not representing the user perspective.
In this paper, we describe our experience of implementing and deploying \textit{ConcepT} in Tencent QQ Browser.
It discovers user-centered concepts at the right granularity conforming to user interests, by mining a large amount of user queries and interactive search click logs. The extracted concepts have the proper granularity, are consistent with user language styles and are dynamically updated.
We further present our techniques to tag documents with user-centered concepts and to construct a \textit{topic-concept-instance} taxonomy, which has helped to improve search as well as news feeds recommendation in Tencent QQ Browser. We performed extensive offline evaluation to demonstrate that our approach could extract concepts of higher quality compared to several other existing methods.
Our system has been deployed in Tencent QQ Browser. Results from online A/B testing involving a large number of real users suggest that the Impression Efficiency of feeds users increased by 6.01\% after incorporating the user-centered concepts into the recommendation framework of Tencent QQ Browser.
\end{abstract}

%
% The code below is generated by the tool at http://dl.acm.org/ccs.cfm.
% Please copy and paste the code instead of the example below.
%


%
% Keywords. The author(s) should pick words that accurately describe the work being
% presented. Separate the keywords with commas.

% % 
% % A "teaser" image appears between the author and affiliation information and the body 
% % of the document, and typically spans the page. 
% \begin{teaserfigure}
%   \includegraphics[width=\textwidth]{sampleteaser}
%   \caption{Seattle Mariners at Spring Training, 2010.}
%   \Description{Enjoying the baseball game from the third-base seats. Ichiro Suzuki preparing to bat.}
%   \label{fig:teaser}
% \end{teaserfigure}

%
% This command processes the author and affiliation and title information and builds
% the first part of the formatted document.
\maketitle
%!TEX root = main.tex
\section{Introduction}
\label{sec:intro}

The capability of \emph{conceptualization} is a critical ability in natural language understanding and is an important distinguishing factor that separates a human being from the current dominating machine intelligence based on vectorization. For example, by observing the words ``Honda Civic'' and ``Hyundai Elantra'', a human can immediately link them with ``fuel-efficient cars'' or ``economy cars'', and quickly come up with similar items like ``Nissan Versa'' and probably ``Ford Focus''. When one observes the seemingly uncorrelated words ``beer'', ``diaper'' and ``Walmart'', one can extrapolate that the article is most likely discussing topics like marketing, business intelligence or even data science, instead of talking about the actual department store ``Walmart''. The importance of concepts is best emphasized by the statement in Gregory Murphy's famous book \emph{The Big Book of Concepts} that ``Concepts embody our knowledge of the kinds of things there are in the world. Tying our past experiences to our present interactions with the environment, they enable us to recognize and understand new objects and events.'' %Quotes from Gregory Murphy have also been cited by Probase \cite{wu2012probase}, a prior study on concept mining conducted by Microsoft.  

%Concepts are mental representations of worldly facts that make up the fundamental building blocks of thoughts and beliefs. They play an important role in all aspects of cognition. To enable machines to better understand natural language, we need to enable them to learn concepts from text. Therefore, mining concepts from text and identifying the relationship between them are fundamental tasks for applications such as query understanding in search engine, document understanding, knowledge base or taxonomy construction, information extraction and so on. Specifically, our objective here is mining concepts to help understanding user queries and documents.

In order to enable machines to extract concepts from text, a large amount of effort has been devoted to knowledge base or taxonomy construction, typically represented by DBPedia \cite{lehmann2015dbpedia} and YAGO \cite{suchanek2007yago} which construct taxonomies from Wikipedia categories, and Probase \cite{wu2012probase} which extracts concepts from free text in web documents.
However, we argue that these methods for concept extraction and taxonomy construction are still limited as compared to how a human interacts with the world and learns to conceptualize, and may not possess the proper granularity that represents human interests.  
%best-suited for understanding the implicit intention in user queries, as well as characterizing the topics of documents.
For example, ``Toyota 4Runner'' is a ``Toyota SUV'' and ``F150'' is a ``truck''. However, it would be more helpful if we can infer that a user searching for these items may be more interested in ``cars with high chassis'' or ``off-road ability'' rather than another Toyota SUV like ``RAV4''---these concepts are rare in existing knowledge bases or taxonomies.
Similarly, if an article talks about the movies ``the Great Gatsby'', ``Wuthering Heights'' and ``Jane Eyre'', it is also hard to infer that the article is actually about ``book-to-film adaptations''. The fundamental reason is that 
taxonomies such as DBPedia \cite{lehmann2015dbpedia} and Probase \cite{wu2012probase}, although maintaining structured knowledge about the world, are not designed to conceptualize from the \emph{user's perspective} or to infer the user intention. Neither can they exhaust all the complex connections between different instances, concepts and topics that are discussed in different documents. Undoubtedly, the ability for machines to conceptualize just as a user would do---to extract trending and user-centered concept terms that are constantly evolving and are expressed in user language---is critical to boosting the intelligence of recommender systems and search engines.

 %(we put Chinese translation in parenthesis because our work are currently done based on Chinese data). 

%However, such concepts rarely appear in web pages in exactly the same form. Therefore, they cannot be extracted using existing approaches.
In this paper, we propose \textit{ConcepT}, a concept mining system at Tencent that aims to discover concepts at the right granularity conforming to user interests.  
Different from prior work, \textit{ConcepT} is not based on mining web pages only, but mining from huge amounts of query logs and search click graphs, thus being able to understand user intention by capturing their interaction with the content.
We present our design of \textit{ConcepT} and our experience of deploying it in Tencent QQ Browser, which has the largest market share in Chinese mobile browser market with more than 110 millions daily active users.
% It provides a content aggregation and feeds recommendation service in its app front page and also powers content loading in Tencent WeChat which has more than 1 billion daily active users all over the world.
\emph{ConcepT} serves as the core taxonomy system in Tencent QQ Browser to discover both time-invariant and trending concepts.
% from query logs.

\emph{ConcepT} can significantly boost the performance of both searching and content recommendation, through the taxonomy constructed from the discovered user-centered concepts as well as a concept tagging mechanism for both short queries and long documents that accurately depict user intention and document coverage.
%First, it extracts featured, user-centered and time-sensitive concepts based on large-scale real world user queries and query logs.
%Second, it is able to associate concepts to documents to help better understanding the topic of documents and improving the performance of both searching and recommendation.
%In addition, \emph{ConcepT} also identifies the \textit{isA} relationship between topics, concepts and instances to form a layered \textit{topic-concept-instance} taxonomy, which together with concept extraction and document tagging significantly improves the performance of both searching and recommendation.
Up to today, \emph{ConcepT} has extracted more than $200,000$ high-quality user-centered concepts from daily query logs and user click graphs in QQ Browser, while still growing at a rate of $11,000$ new concepts found per day. Although our system is implemented and deployed for processing Chinese query and documents, the proposed techniques in \emph{ConcepT} can easily be adapted to other languages.

% First, existing knowledge bases or taxonomies, such as DBPedia \cite{lehmann2015dbpedia}, YAGO \cite{suchanek2007yago}, and Probase \cite{wu2012probase}, mainly consist of \textit{coarse} level general concepts and \textit{fine} level instances. However, many medium level concepts are missing. We may have instances ``China'', ``India'' in the concepts ``countries'' and ``developing countries''. But we cannot recognize that they are also ``populous Asian countries'' or ``countries with ancient civilization'' based on limited concept space. Therefore, mining such kind of fine-grained and featured concepts helps with a better understanding of queries and documents.

Mining user-centered concepts from query logs and search click graphs has brought about a number of new challenges.
First, most existing taxonomy construction approaches such as Probase \cite{wu2012probase}
extract concepts based on Hearst patterns, like ``such as'', ``especially'', etc. However, Hearst patterns have limited extraction power, since high-quality patterns are often missing in short text like queries and informal user language.
Moreover, existing methods extract concepts from web pages and documents that are usually written by experts %In this way, the extracted concepts tend to be in very formal expression. I 
in the \emph{writer perspective}. However, search queries are often informal and may not observe the syntax of a written language \cite{hua2015short}. Hence, it is hard if not impossible to mine ``user perspective'' concepts based on predefined syntax patterns.

There are also many studies on keyphrase extraction \cite{shang2018automated,liu2015mining,mihalcea2004textrank}. They measure the importance or quality of all the $N$-grams in a document or text corpus, and choose keyphrases from them according to the calculated scores. As a result, such methods can only extract continuous text chunks, whereas a concept may be discontinuous or may not even be explicitly mentioned in a query or a document. Another concern is that most of such $N$-gram keyphrase extraction algorithms yield poor performance on short text snippets such as queries. In addition, deep learning models, such as sequence-to-sequence, can also be used to generate or extract concepts. However, deep learning models usually rely on large amounts of high-quality training data. For user-centered concept mining, manually labeling such a dataset from scratch is extremely costly and time consuming.

Furthermore, many concepts in user queries are related to recent trending events
whereas the concepts in existing taxonomies are mostly stable and time-invariant. A user may search for ``Films for New Year (\begin{CJK}{UTF8}{gkai}贺岁大片\end{CJK})'' or ``New Japanese Animation in April (\begin{CJK}{UTF8}{gkai}四月新番\end{CJK})'' in Tencent QQ Browser. The semantics of such concepts are evolving over time, since apparently we have different new animations or films in different years. Therefore, in contrast to existing taxonomies which mostly maintain long-term stable knowledge, it will be challenging yet beneficial if we can also extract time-varying concepts and dynamically update the taxonomies constructed.

We make the following novel contributions in the design of \emph{ConcepT}:

\emph{First}, we extract candidate user-centered concepts from vast query logs by two unsupervised strategies: 1) bootstrapping based on pattern-concept duality: a small number of predefined string patterns can be used to find new concepts while the found concepts can in turn be used to expand the pool of such patterns; 2) query-title alignment: an important concept in a query would repeat itself in the document title clicked by the user that has input the query. 

\emph{Second}, we further train a supervised sequence labeling Conditional Random Field (CRF) and a discriminator based on the initial seed concept set obtained, to generalize concept extraction and control the concept quality.
These methods are complementary to each other and are best suited for different cases. Evaluation based on a labeled test dataset has shown that our proposed concept discovery procedure significantly outperforms a number of existing schemes.

%Compared with existing concept mining approaches, the expression of concepts extracted by our system are consistent with users' language style, i.e., they are expressed in the manner of ``user perspective'' rather than ``writer perspective''. The semantic granularity of concepts naturally matches user intention, and the concepts are also time-sensitive. That is because the user queries and click logs, where the concepts are extracted from, are time-sensitive and reveals user intention. 

%Based on it, on one hand, we can rewrite user queries by query conceptualization, including instantiation that replace a concept with its typical and likely instances, and abstraction that replace one or multiple instances with the typical concepts they belong to \cite{wu2012probase}. 

%In this way, we can improve the search results of search engines. On another hand, with better understanding of document topics, we can recommend them to appropriate users and improve recommendation performance.
\emph{Third}, we propose effective strategies to tag documents with potentially complex concepts to depict document coverage, mainly by combining two methods: 1) matching key instances in a document with their concepts if their \emph{isA} relationships exist in the corresponding constructed taxonomy; 2) using a probabilistic inference framework to estimate the probability of a concept provided that an instance is observed in its context. Note that the second method can handle the case when the concept words do not even appear in the document.
For example, we may associate an article containing ``celery'', ``whole wheat bread'' and ``tomato'' with the concept ``diet for weight loss'' that a lot of users are interested in, even if the document does not have exact wording for ``weight loss'' but has context words such as ``fibre'', ``healthy'', and ``hydrated''.

\emph{Last but not least}, we have constructed and maintained a three-layered \textit{topic-concept-instance} taxonomy, by identifying the \emph{isA} relationships among instances, concepts and topics based on machine learning methods, including deep neural networks and probabilistic models.
Such a user-centered taxonomy significantly helps with query and document understanding at varying granularities.

%By associating a document with matching concepts that may never even be mentioned in the document, we can better characterize the topic of documents from the user perspective and improve search engines by query conceptualization or instantiation. %We will introduce more details in Sec.~\ref{sec:modeldoc}. 

We have evaluated the performance of \emph{ConcepT}, and observed that it can improve both searching and recommendation results, through both offline experiments and a large-scale online A/B test on more than $800,000$ real users conducted in the QQ Browser mobile app.
The experimental results reveal that our proposed methods can extract concepts more accurately from Internet user queries in contrast to a variety of existing approaches. 
Moreover, by performing query conceptualization based on the extracted concepts and the correspondingly constructed taxonomy, we can improve the results of search engine according to a pilot user experience study in our experiments. 
Finally, \emph{ConcepT} also leads to a higher Impression Efficiency as well as user duration in the real world according to the large-scale online A/B test on the recommender system in feeds stream (text digest content recommended to users in a stream as they scroll down in the mobile app). 
The results suggest that the Impression Efficiency of the users increases by 6.01\% when \emph{ConcepT} system is incorporated for feeds stream recommendation.

%!TEX root = main.tex
\section{User-Centered Concept Mining}
\label{sec:modelquery}

\begin{figure*}[tb]
\centering
\includegraphics[width=0.9\textwidth]{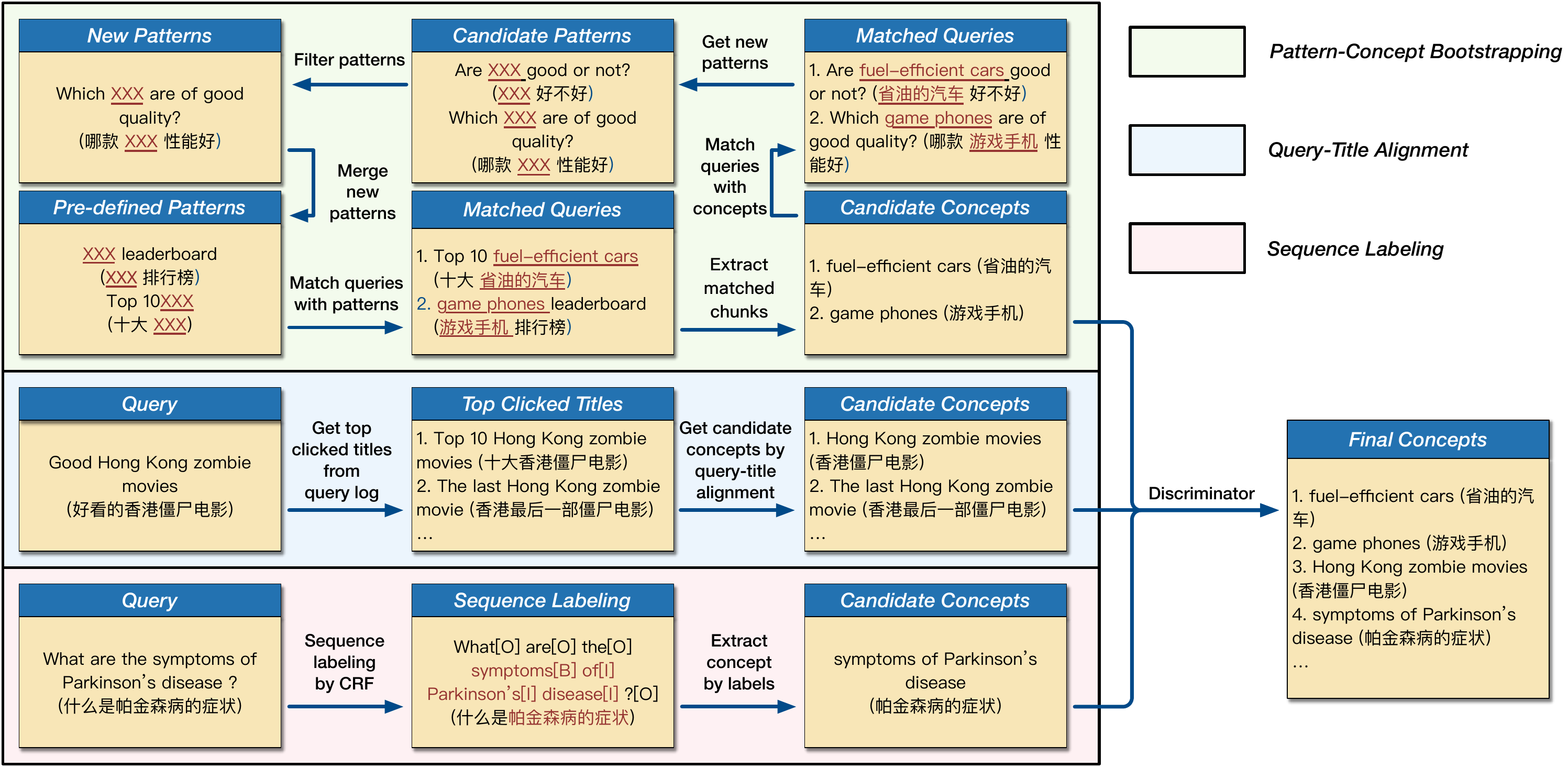}
\vspace{-3mm}
\caption{The overall process of concept mining from user queries and query logs.}
\label{fig:QueryConcept}
\vspace{-3mm}
\end{figure*}

Our objective of user-centered concept mining is to derive a word/phrase from a given user query which can best characterize this query and its related click logs at the proper granularity. 

Denote a user query by $q = w_1^qw_2^q\cdots w_{|q|}^q$, which is a sequence of words. Let $Q$ be the set of all queries. Denote a document title by $t = w_1^tw_2^t\cdots w_{|t|}^t$, another sequence of words. 
Given a user query $q$ and its corresponding top-ranked clicked titles $T^q = \{t^q_1, t^q_2, \cdots, t^q_{|T^q|}\}$ from query logs, we aim to extract a concept phrase $\mathbf{c} = w^c_1w^c_2\cdots w^c_{|\mathbf{c}|}$ that represents the main semantics or the intention of the query. Each word $w^c_i \in \mathbf{c}$ belongs to either the query $q$ or one of the corresponding clicked titles $t^q_j \in T^q$.

An overview of the detailed steps of user-centered concept mining from queries and query logs in \emph{ConcepT} is shown in Fig.~\ref{fig:QueryConcept}, which mainly consists of three approaches: pattern-concept bootstrapping, query-title alignment, as well as supervised sequence labeling. All the extracted concepts are further filtered by a discriminator. We utilize bootstrapping and query-title alignment to automatically accumulate an initial seed set of \textit{query-concept} pairs, which can help to train sequence labeling and the discriminator, to extract a larger amount of concepts more accurately.

\textbf{Bootstrapping by Pattern-Concept Duality}.
We first extract an initial set of seed concepts by applying the bootstrapping idea \cite{brin1998extracting} only to the set of user queries $Q$ (without the clicked titles). Bootstrapping exploits \textit{Pattern-Concept Duality}, which is:
\begin{itemize}
	\item Given a set of patterns, we can extract a set of concepts from queries following these patterns.
	\item Given a set of queries with extracted concepts, we can learn a set of patterns. 
\end{itemize}
% basis of seed concepts mining in our system.
Fig.~\ref{fig:QueryConcept} (a) illustrates how bootstrapping is performed on queries $Q$. First, we manually define a small set of patterns which can be used to accurately extract concept phrases from queries with high confidence. For example, %``\underline{XXX} leaderboard (\begin{CJK}{UTF8}{gkai}\underline{XXX}排行榜\end{CJK})'', 
``Top 10 \underline{XXX} (\begin{CJK}{UTF8}{gkai}十大\underline{XXX}\end{CJK})'' is a pattern (with original Chinese expression in parenthesis) that can be used to extract seed concepts. %``XXX'' denotes any text chunk. 
Based on this pattern, we can extract concepts: ``fuel-efficient cars (\begin{CJK}{UTF8}{gkai}省油的汽车\end{CJK})'' and ``gaming phones (\begin{CJK}{UTF8}{gkai}游戏手机\end{CJK})'' from the queries ``Top 10 \underline{fuel-efficient cars} (\begin{CJK}{UTF8}{gkai}十大\underline{省油的汽车}\end{CJK})'' and ``Top 10 \underline{gaming phones} (\begin{CJK}{UTF8}{gkai}十大\underline{游戏手机}\end{CJK})'', respectively. %The extracted query-concept pairs become samples in our training dataset.

We can in turn retrieve more queries that contain these extracted concepts and derive new patterns from these queries. For example, a query ``Which gaming phones have the best performance? (\begin{CJK}{UTF8}{gkai}哪款游戏手机性能好?\end{CJK})'' also contains the concept ``gaming phones (\begin{CJK}{UTF8}{gkai}游戏手机\end{CJK})''. Based on this query, a new pattern ``Which \underline{XXX} have the best performance? (\begin{CJK}{UTF8}{gkai}哪款\underline{XXX}性能好?\end{CJK})'' is found.

We also need to shortlist and control the quality of the patterns found in each round.
%$For each extracted seed concept $\mathbf{c}$, denote the set of queries containing it as $Q_{c}$. We maintain $Q_{c}$ for each concept during bootstrapping.
Intuitively speaking, a pattern is valuable if it can be used to accurately extract a portion of existing concepts as well as to discover new concepts from queries. However, if the pattern is too general and appears in a lot of queries, it may introduce noise. For example, a pattern ``Is \underline{XXX} good? (\begin{CJK}{UTF8}{gkai}\underline{XXX}好不好?\end{CJK})'' underlies a lot of queries including ``Is \underline{the fuel-efficient car} good? (\begin{CJK}{UTF8}{gkai}\underline{省油的车}好不好?\end{CJK})'' and ``Is \underline{running everyday} good (\begin{CJK}{UTF8}{gkai}\underline{每天跑步}好不好?\end{CJK})'', whereas ``running everyday (\begin{CJK}{UTF8}{gkai}每天跑步\end{CJK})'' does not serve as a sufficiently important concept in our system. 
Therefore, given a new pattern $\mathbf{p}$ found in a certain round, let $n_{s}$ be the number of concepts in the existing seed concept set that can be extracted from query set $Q$ by $\mathbf p$. Let $n_e$ be the number of new concepts that can be extracted by $\mathbf p$ from $Q$. We will keep the pattern $\mathbf{p}$ if it satisfies: 1) $\alpha < \frac{n_s}{n_e} < \beta$, and 2) $n_s > \delta$, where $\alpha, \beta, \text{and } \delta$ are predefined thresholds. (We set $\alpha = 0.6,\ \beta=0.8,\ \text{and} \ \delta=2$ in our system.)

 %In the following, we further propose a query-title alignment strategy and sequence labeling-based method to improve the recall rate of concept mining.

\textbf{Concept mining by query-title alignment}.
Although bootstrapping helps to discover new patterns and concepts from the query set $Q$ in an iterative manner, such a pattern-based method has limited extraction power. Since there are a limited number of high-quality syntax patterns in queries, the recall rate of concept extraction has been sacrificed for precision.
Therefore, we further propose to extract concepts from both a query and its top clicked link titles in the query log.

The intuition is that a concept in a query will also be mentioned in the clicked titles associated with the query, yet possibly in a more detailed manner.
For example,  ``The last Hong Kong zombie movie (\begin{CJK}{UTF8}{gkai}香港|最后|一|部|僵尸|电影\end{CJK})'' or ``Hong Kong zombie comedy movie (\begin{CJK}{UTF8}{gkai}香港|搞笑|僵尸|电影\end{CJK})'' convey more specific concepts of the query ``Hong Kong zombie movie (\begin{CJK}{UTF8}{gkai}香港|僵尸|电影\end{CJK})'' that leads to the click of these titles. Therefore, we propose to find such concepts based on the alignment of queries with their corresponding clicked titles. The steps are listed in the following:
\begin{enumerate}
\item Given a query $q$, we retrieve the top clicked titles $T^q = \{t^q_1, t^q_2, \cdots, t^q_{|T^q|}\}$ from the query logs of $q$, i.e., $T^q$ consists of document titles that are clicked by users for more than $N$ times during the past $D$ days ($N=5$ and $D=30$ in our system).

\item For query $q$ and each title $t \in T^q$, we enumerate all the $N$-grams in them.

\item Let $N$-gram $g^q_{in} = w^q_i w^q_{i+1} \cdots w^q_{i + n - 1}$ denote a text chunk of length $n$ starting from position $i$ of query $q$,  and \\ $g^t_{jm} = w^t_j w^t_{j+1} \cdots w^t_{j + m - 1}$ denote a text chunk of length $m$ starting from position $j$ of title $t$.
For each pair of such $N$-grams, $<g^q_{in}, g^t_{jm}>$, we identify $g^t_{jm}$ as a candidate concept if: i) $g^t_{jm}$ contains all the words of $g^q_{in}$ in the same order; ii) $w^q_i = w^t_j$, and $w^q_{i + n - 1} = w^t_{j + m - 1}$.
\end{enumerate}
Query-title alignment extends concept extraction from query set alone to concept discovery based on the query logs, thus incorporating some information of the user's interaction into the system. 

\textbf{Supervised sequence labeling}. The above unsupervised methods are still limited in their generalizability. We further perform supervised learning and train a Conditional Random Field (CRF) to label the sequence of concept words in a query or a title, where the training dataset stems from the results of the bootstrapping and query-title alignment process mentioned above, combined with human reviews as detailed in the Appendix.
Specifically, each word is represented by its tag features, e.g., Part-of-Speech or Named Entity Recognition tags, and the contextual features, e.g., the tag features of its previous word and succeeding word, the combination pattern of tags of contextual words and the word itself. These features are fed into a CRF to yield a sequence of labels, identifying the concept chunk, as shown in Fig.~\ref{fig:QueryConcept}.

The above approaches for concept mining are complementary to each other. Our experience shows that CRF can better extract concepts from short text when they have clear boundary with surrounding non-concept words, e.g., ``What cars are fuel-efficient (\begin{CJK}{UTF8}{gkai}\underline{省油的汽车}有哪些\end{CJK})''. However, when the concept is split into multiple pieces, e.g., ``What \underline{gifts} should we prepare \underline{for birthdays of parents}? (\begin{CJK}{UTF8}{gkai}\underline{父母}过\underline{生日}准备什么\underline{礼物}?\end{CJK})'', the query-title alignment approach can better capture the concept that is scattered in a query.

\textbf{A Discriminator for quality control.}
Given the concepts extracted by above various strategies, we need to evaluate their value. For example, in Fig.~\ref{fig:QueryConcept}, the concept ``The last Hong Kong zombie movie (\begin{CJK}{UTF8}{gkai}香港|最后|一|部|僵尸|电影\end{CJK})'' is too fine-grained and maybe only a small amount of users are interested in searching it. Therefore, we further train a classifier to determine whether each discovered concept is worth keeping.

We represent each candidate concept by a variety of its features such as whether this concept has ever appeared as a query, how many times it has been searched and so on (more details in Appendix). With these features serving as the input, we train a classifier, combining Gradient Boosting Decision Tree (GBDT) and Logistic Regression, to decide whether to accept the candidate concept in the final list or not. The training dataset for the discriminator is manually created. We manually check a found concept to see whether it is good (positive) or not sufficiently good (negative). %If yes, we label it as positive sample, otherwise it is a negative one. 
Our experience reveals that we only need $300$ samples to train such a discriminator. Therefore, the creation of the dataset incurs minimum overhead.

%!TEX root = main.tex
\section{Document Tagging and Taxonomy Construction}
\label{sec:modeldoc}

\begin{figure}[tb]
\centering
\includegraphics[width=2.7in]{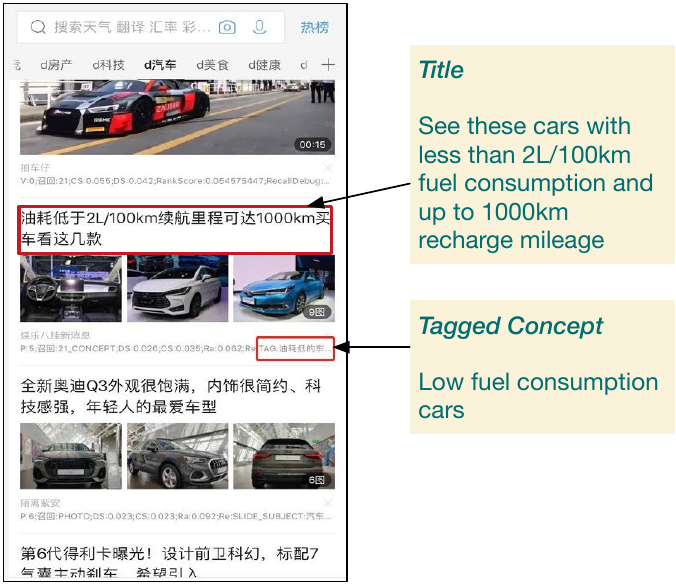}
\vspace{-3mm}
\caption{Example of concept tagging for documents in the feeds stream of Tencent QQ Browser.}
\label{fig:DocTagExample}
\vspace{-3mm}
\end{figure}

In this section, we describe our strategies for tagging each  document with pertinent user-centered concepts to depict its coverage. Based on document tagging, we further construct a 3-layered \textit{topic-concept-instance} taxonomy which helps with feeds recommendation in Tencent QQ Browser.

\subsection{Concept Tagging for Documents}

\begin{figure*}[tb]
\centering
\includegraphics[width=0.9\textwidth]{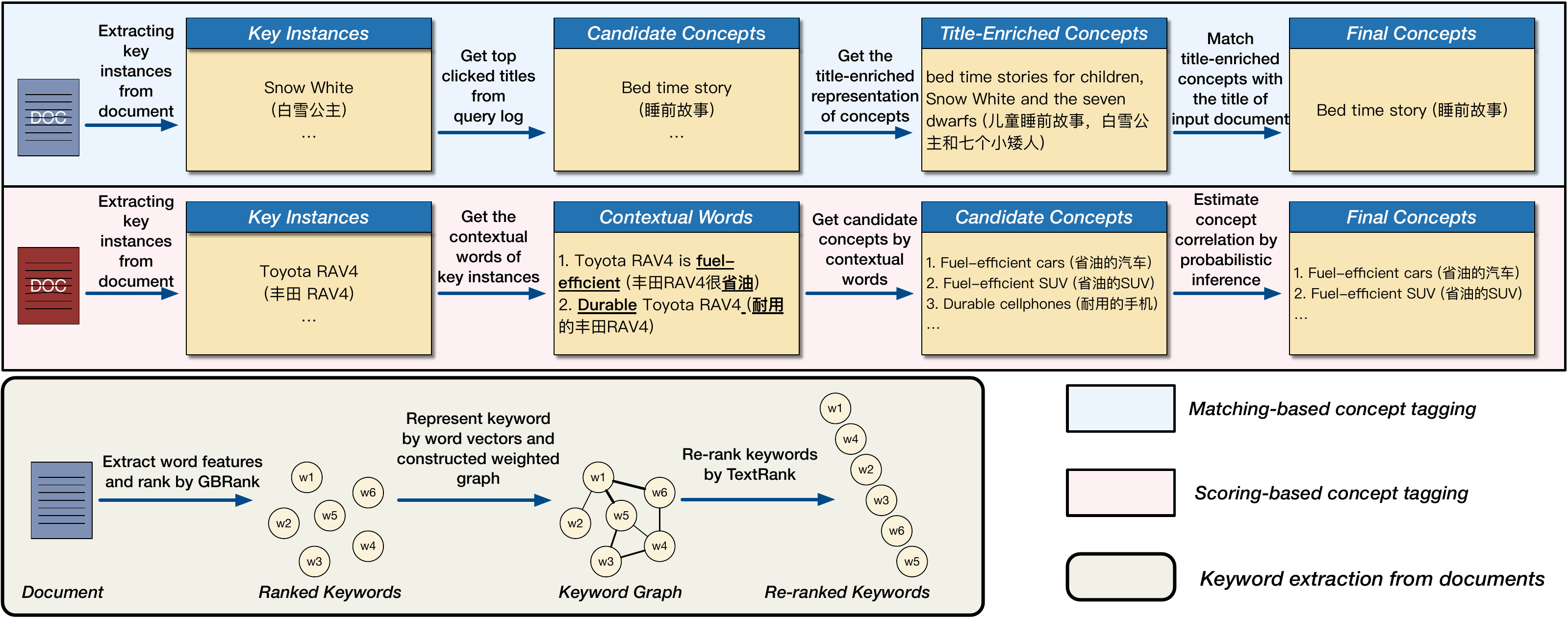}
\vspace{-3mm}
\caption{The overall procedures of concept tagging for documents. We combine both a matching-based approach with a scoring-based approach to handle different situations.}
\label{fig:DocConcept}
\vspace{-3mm}
\end{figure*}

While the extracted concepts can characterize the implicit intention of user queries, they can also be used to describe the main topics of a document.
Fig.~\ref {fig:DocTagExample} shows an example of concept tagging in Tencent QQ Browser based on the \emph{ConcepT} system. Suppose that a document titled ``See these cars with less than 2L/100km fuel consumption and up to 1000km recharge mileage'' can be tagged with the concept ``low fuel-consumption cars'', even though the title never explicitly mentions these concept words. Such concept tags for documents, if available, can help improve search and recommendation performance.
Therefore, we propose to perform concept tagging for documents.

Given a document $d$ and a set of concepts $\mathbf{C} = \{\mathbf{c}_1, \mathbf{c}_2, \cdots, \mathbf{c}_{|\mathbf{C}|}\}$, our problem is selecting a subset of concepts $\mathbf{C}^d = \{\mathbf{c}^d_1, \mathbf{c}^d_2, \cdots, \mathbf{c}^d_{|\mathbf{C}^d|}\}$ from $\mathbf{C}$ that are most related to the content of $d$.
Fig.~\ref{fig:DocConcept} presents the procedures of concept tagging for documents.
To link appropriate concepts with a document, we propose a probabilistic inference-based approach, together with a matching-based method to handle different situations.

Specifically, our approach estimates the correlation between a concept and a document through the key instances contained in the document.
When no direct \textit{isA} relationship can be found between the key instances in the document and the concepts, we use probabilistic inference as a general approach to identify relevant concepts by utilizing the context information of the instances in the document.
Otherwise, the matching-based approach retrieves candidate concepts which have \textit{isA} relationships with the key instances in a taxonomy we have constructed (which be explained at the end of this section). After that, it scores the coherence between a concept and a document based on the title-enriched representation of the concept.

\textbf{Key instance extraction.}
Fig.~\ref{fig:DocConcept} shows our approach for key instance extraction. Firstly, we rank document words using GBRank \cite{zheng2007regression} algorithm, based on  word frequency, POS tag, NER tag, etc. Secondly, we represent each word by word vectors proposed in \cite{N18-2028}, and construct a weighted undirected graph for top $K$ words (we set $K=10$). The edge weight is calculated by the cosine similarity of two word vectors. We then re-rank the keywords by TextRank \cite{mihalcea2004textrank} algorithm.  Finally, we only keep keywords with ranking scores larger than $\delta_w$ (we use 0.5).
From our experience, combining GBRank and word-vector-based TextRank helps to extract keywords that are more coherent to the topic of document.

\textbf{Concept tagging by probabilistic inference.}
%The matching-based approach for concept tagging requires the established \textit{isA} relationship between instances and concepts. However, we may miss a certain amount of relationship in our taxonomy. In this case, we instead detect the correlated concepts through a probabilistic inference-based approach.
Denote the probability that concept $\mathbf{c}$ is related to document $d$ as $p(\mathbf{c}|d)$. We propose to estimate it by:
\begin{align}
p(\mathbf{c}|d) &= \sum_{i=1}^{|E^d|}p(\mathbf{c}|\mathbf{e}^d_i) p(\mathbf{e}^d_i | d),
\end{align}
where $E^d$ is the key instance set of $d$, and $p(\mathbf{e}^d_i | d)$ is the document frequency of instance $\mathbf{e}^d_i \in E^d$.
$p(\mathbf{c}|\mathbf{e}^d_i)$ estimates the probability of concept $\mathbf{c}$ given $\mathbf{e}^d_i$. However, as the \textit{isA} relationship between $\mathbf{e}^d_i$ and $\mathbf{c}$ may be missing, we further infer the conditional probability by taking the contextual words of $\mathbf{e}^d_i$ into account:
\begin{align}
\label{eq:pce}
p(\mathbf{c}|\mathbf{e}^d_i) = \sum_{j = 1}^{|X_{E^d}|} p(\mathbf{c} | \mathbf{x}_j) p(\mathbf{x}_j | \mathbf{e}^d_i)
\end{align}
$p(\mathbf{x}_j | \mathbf{e}^d_i)$ is the co-occurrence probability of context word $\mathbf{x}_j$ with $e^d_i$. We consider two words as co-occurred if they are contained in the same sentence. $X_{E^d}$ are the set of contextual words of $\mathbf{e}^d_i$ in $d$. Denote $\mathbf{C}^{\mathbf{x}_j}$ as the set of concepts containing $\mathbf{x}_j$ as a substring.
$p(\mathbf{c} | \mathbf{x}_j)$ is defined as:
\begin{align} 
p(\mathbf{c} | \mathbf{x}_j) = 
\begin{cases}\frac{1}{|\mathbf{C}^{\mathbf{x}_j}|}. & \text{if } \mathbf{x}_j \text{is a substring of } \mathbf{c},\\
0 & \text{otherwise.}
\end{cases}
\end{align}
For example, in Fig.~\ref{fig:DocConcept}, suppose $\mathbf{e}^d_i$ extracted from $d$ is ``Toyota RAV4 (\begin{CJK}{UTF8}{gkai}丰田RAV4\end{CJK})''. We may haven't establish any relationship between this instance and any concept. However, we can extract contextual words ``fuel-efficient (\begin{CJK}{UTF8}{gkai}省油\end{CJK})'' and ``durable (\begin{CJK}{UTF8}{gkai}耐用\end{CJK})'' from $d$. Based on these contextual words, we can retrieve candidate concepts that containing these words, such as ``fuel-efficient cars (\begin{CJK}{UTF8}{gkai}省油的汽车\end{CJK})'' and ``durable cellphones (\begin{CJK}{UTF8}{gkai}耐用的手机\end{CJK})''. We then estimate the probability of each candidate concept by above equations.

\textbf{Concept tagging by matching.}
The probabilistic inference-based approach decomposes the correlation between a concept and a document through the key instances and their contextual words in the document.
However, whenever the \textit{isA} relationship between the key instances of $d$ and $\mathbf{C}$ is available, we can utilize it to get candidate concepts directly, and calculate the matching score between each candidate concept and $d$ to decide which concepts are coherent to the document.

First, we introduce how the \textit{isA} relationship between \textit{concept-instance} pairs can be identified.
On one hand, given a concept, we retrieve queries/titles containing the same modifier in the context and extract the instances contained in the queries/titles. For example, given concept ``fuel-efficient cars (\begin{CJK}{UTF8}{gkai}省油的汽车\end{CJK})'', we may retrieve a query/title ``fuel-efficient Toyota RAV4 (\begin{CJK}{UTF8}{gkai}省油的丰田RAV4\end{CJK})'', and extract instance ``Toyota RAV4 (\begin{CJK}{UTF8}{gkai}丰田RAV4\end{CJK})'' from the query/title, as it shares the same modifier ``fuel-efficient (\begin{CJK}{UTF8}{gkai}省油的\end{CJK})'' with the given concept.
After we getting a candidate instance $\mathbf{e}$, we estimate $p(\mathbf{c} | \mathbf{e})$ by Eqn. (\ref{eq:pce}).
On another hand, we can also extract \textit{concept-instance} pairs from various semi-structured websites where a lot of \textit{concept-instance} pairs are stored in web tables.

Second, we describe our matching-based approach for concept tagging.
Let $E^d = \{\mathbf{e}^d_1, \mathbf{e}^d_2, \cdots, \mathbf{e}^d_{|E^d|}\}$ donate the set of key instances extracted from $d$, and $C^d = \{\mathbf{c}^d_1, \mathbf{c}^d_2, \cdots, \mathbf{c}^d_{|C^d|}\}$ donate the retrieved candidate concepts by the \textit{isA} relationship of instances in $E^d$.
For each candidate concept $\mathbf{c}^d_i$, we enrich its representation by concatenating the concept itself with the top $N$ (we use 5) titles of user clicked links.
% We then combine two strategies for concept-document matching:
% \begin{itemize}
% 	\item Represent enriched concept and document by TF-IDF vector, and calculate cosine similarity. If the similarity $sim(\mathbf{c}, d)$. If is above $sim(\mathbf{c}, d) > \red{\delta_u}$, we tag $\mathbf{c}$ to $d$; if $sim(\mathbf{c}, d) < \red{\delta_l}$, we reject it.
% 	\item If $\red{\delta_l} \leq sim(\mathbf{c}, d) \leq \red{\delta_u}$, we match $\mathbf{c}$ and $d$ by Siamese Encoded Graph Convolutional Network (SE-GCN) \cite{liu2018matching}, which encodes text documents by graph convolution and matches two documents through Siamese structured neural network. More details about SE-GCN can be found in \cite{liu2018matching}.
% \end{itemize}
We then represent enriched concept and the document title by TF-IDF vectors, and calculate the cosine similarity between them. If $sim(\mathbf{c}, d) > \delta_u$ (we set it as 0.58), we tag $\mathbf{c}$ to $d$; otherwise we reject it.
Note that the \textit{isA} relationship between \textit{concept-instance} pairs and the enriched representation of concepts are all created in advance and stored in a database.

Fig.~\ref{fig:DocConcept} shows an example of matching-based concept tagging. Suppose we extract key instance ``Snow White (\begin{CJK}{UTF8}{gkai}白雪公主\end{CJK})'' from a document, we can retrieve related concepts ``bed time stories (\begin{CJK}{UTF8}{gkai}睡前故事\end{CJK})'' and ``fairy tales (\begin{CJK}{UTF8}{gkai}童话故事\end{CJK})'' based on \textit{isA} relationship. The two concepts are further enriched by the concatenation of top clicked titles. Finally, we match candidate concepts with the original document, and keep highly related concepts.

\subsection{Taxonomy Construction}
\label{subsec:taxonomy}

\begin{figure}[tb]
\centering
\includegraphics[width=0.475\textwidth]{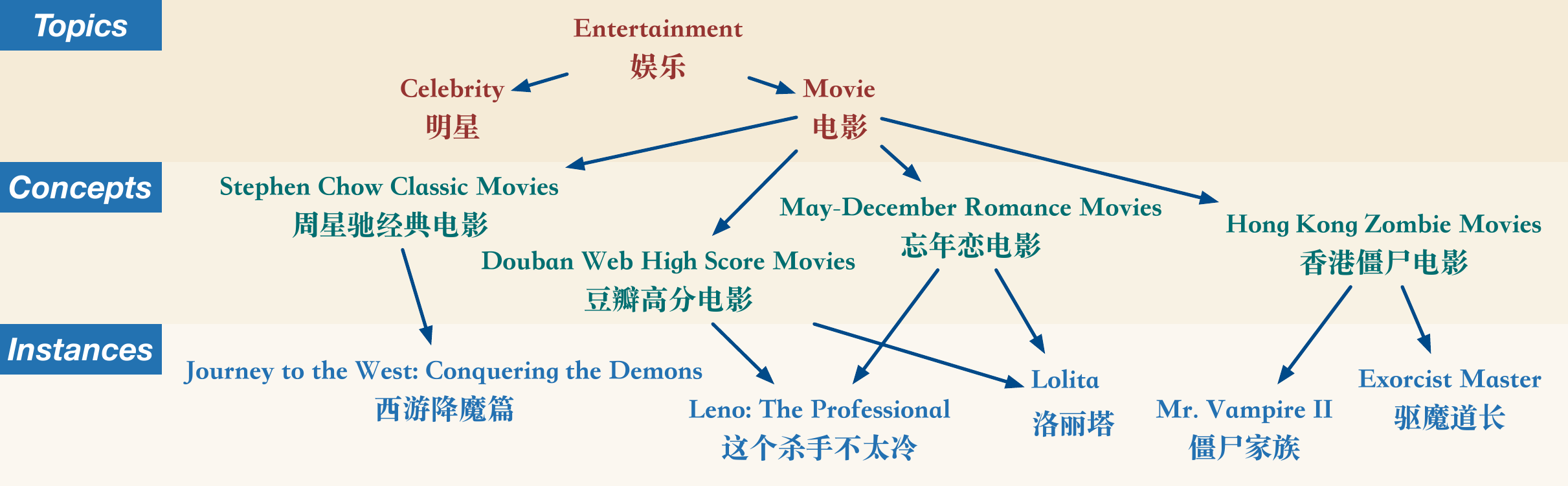}
\vspace{-5mm}
\caption{An example to show the extracted topic-concept-instance hierarchy.}
\label{fig:CaseStudy}
\vspace{-5mm}
\end{figure}

% Aside from concept tagging for documents, we further introduce our algorithms for keyword extraction and topic classification. 
% Based on the topics, concepts, and key instances of documents,

We have also constructed a \textit{topic-concept-instance} taxonomy based on the concepts extracted from queries and query logs. It can reveal the hierarchical relationship between different topics, concepts and instances.
Currently our constructed taxonomy consists of $31$ pre-defined topics, more than $200,000$ user-centered concepts, and more than $600,000$ instances.
Among them, 40,000 concepts contain at least one instance, and 200,000 instances have been identified with a \emph{isA} relationship with at least one concept.
Based on the taxonomy, we can improve the user experience in search engines by understanding user implicit intention via query conceptualization, as well as enhance the recommendation performance by matching users and documents at different semantic granularities. We will demonstrate such improvements in detail in Sec.~\ref{sec:eval}.

Fig.~\ref{fig:CaseStudy} shows a three-layered taxonomy that consists of topics, concepts and instances. The taxonomy is a Directed Acyclic Graph (DAG). Each node is either a topic, a concept or an instance. We predefined a list that contains $N_t = 31$ different topics, including entertainment, technology, society and so on. The directed edges indicate \textit{isA} relationships between nodes.

We have already introduced our approach for \textit{isA} relationship between \textit{concept-instance} pairs.
We need to further identify the relationship between \textit{topic-concept} pairs.
First, we represent each document as a vector through word embedding and pooling, and perform topic classification for documents through a carefully designed deep neural network (see Appendix for details).  
After that, given a concept $\mathbf{c}$ and a topic $\mathbf{p}$, suppose there are $n^{\mathbf{c}}$ documents that are correlated to concept $\mathbf{c}$, and among them there are $n^{\mathbf{c}}_{\mathbf{p}}$ documents that belong to topic $\mathbf{p}$. We then estimate $p(\mathbf{p} | \mathbf{c})$ by $p(\mathbf{p} | \mathbf{c}) = n^{\mathbf{c}}_{\mathbf{p}} / n^{\mathbf{c}}$. We identify the \textit{isA} relationship between $\mathbf{c}$ and $\mathbf{p}$ if $p(\mathbf{p} | \mathbf{c}) > \delta_t$ (we set $\delta_t= 0.3$). Our experience shows that most of the concepts belong to one or two topics.

%!TEX root = main.tex
\section{Evaluation}
\label{sec:eval}

In this section, we first introduce a new dataset for the problem of concept mining from user queries and query logs, and compare our proposed approach with variety of baseline methods. We then evaluate the accuracy of the taxonomy constructed from extracted user-centered concepts, and show that it can improve search engine results by query rewriting. Finally, we run large-scale online A/B testing to show that the concept tagging on documents significantly improves the performance of recommendation in real world.

We deploy the \emph{ConcepT} system which includes the capability of concept mining, tagging, and taxonomy construction in Tencent QQ Browser. 
For offline concept mining, our current system is able to extract around 27,000 concepts on a daily basis, where about 11,000 new concepts are new ones.
For online concept tagging, our system processes 40 documents per second.
More details about implementation and deployment can be found in appendix.

\subsection{Evaluation of Concept Mining}

\textbf{The User-Centered Concept Mining Dataset (UCCM).} As user-centered concept mining from queries is a relative new research problem and there is no public dataset available for evaluation, we created a large-scale dataset containing $10,000$ samples.
Our \textit{UCCM} dataset is sampled from the queries and query logs of Tencent QQ Broswer, from November 11, 2017 to July 1, 2018. For each query, we keep the document titles clicked by more than 2 users in previous day.
Each sample consists of a query, the top clicked titles from real world query log, and a concept phrase labeled by $3$ professional product managers in Tencent and $1$ PhD student. We have published the \textit{UCCM} dataset for research purposes \footnote{https://github.com/BangLiu/ConcepT}.
% The remaining titles are further filtered by a binary classifier which detects whether the title contains a concept.
% We trained a logistic regression model where the input features are the bag-of-words feature of $161$ indicator keywords, including ``summary (\begin{CJK}{UTF8}{gkai}汇总\end{CJK})'', ``collection (\begin{CJK}{UTF8}{gkai}集锦\end{CJK})'' and so forth.
%Details about our dataset can be found in the supplementary file.
% \blue{maybe a table about data lengths, statistics.}
%\blue{train, test split, or illustrate that we doesn't need split. Our model is trained previously.}

\textbf{Methodology and Compared Models}. We evaluate our comprehensive concept mining approach with the following baseline methods and variants of our method:
\begin{itemize}
    \item \textbf{TextRank \cite{mihalcea2004textrank}}. The classical graph-based ranking model for keyword extraction.\footnote{https://github.com/letiantian/TextRank4ZH}
    \item \textbf{THUCKE \cite{liu2011automatic}}. It regards keyphrase extraction as a problem of translation, and learns translation probabilities between the words in input text and the words in keyphrases.\footnote{https://github.com/thunlp/THUCKE}
    \item \textbf{AutoPhrase \cite{shang2018automated}}. A state-of-the-art quality phrase mining algorithm that extracts quality phrases based on knowledge base and POS-guided segmentation.\footnote{https://github.com/shangjingbo1226/AutoPhrase}
	\item \textbf{Pattern-based matching with query (Q-Pattern)}. Extract concepts from queries based on patterns from bootstrapping.
  \item \textbf{Pattern-based matching with title (T-Pattern)}. Extract concepts from titles based on patterns from bootstrapping.
	\item \textbf{CRF-based sequence labeling with query (Q-CRF)}. Extract concepts from queries by CRF.
	\item \textbf{CRF-based sequence labeling with titles (T-CRF)}. Extract concepts from click titles by CRF.
	\item \textbf{Query-Title alignment (QT-Align)}. Extract concepts by query-title alignment strategy.
\end{itemize}
For the T-Pattern and T-CRF approach, as each click title will give a result, we select the most common one as the final result given a specific query.
For the TextRank, THUCKE, and AutoPhrase algorithm, we take the concatenation of user query and click titles as input, and extract the top 5 keywords or phrases. We then keep the keywords/phrases contained in the query and concatenate them in the same order as in the query, then use it as the final result.

We use Exact Match (EM) and F1 to evaluate the performance. The exact match score is 1 if the prediction is exactly the same as groundtruth or 0 otherwise.
F1 measures the portion of overlap tokens between the predicted phrase and the groundtruth concept.

\begin{table}[tb]
  \caption{Compare different algorithms for concept mining.}
  \label{tab:concept-mining}
  \begin{tabular}{lll}
    \toprule
    Method & Exact Match & F1 Score\\
    \midrule
    TextRank & $0.1941$ & $0.7356$ \\
    THUCKE & $0.1909$ & $0.7107$ \\
    AutoPhrase & $0.0725$ & $0.4839$ \\
    \midrule
    Q-Pattern & $0.1537$ & $0.3133$ \\
    T-Pattern & $0.2583$ & $0.5046$ \\
    Q-CRF & $0.2631$ & $0.7322$ \\
    T-CRF & $0.3937$ & $0.7892$ \\
    QT-Align & $0.1684$ & $0.3162$ \\
    \midrule
    Our approach & $\mathbf{0.8121}$ & $\mathbf{0.9541}$ \\
    \bottomrule
  \end{tabular}
  \vspace{-5mm}
\end{table}

\textbf{Evaluation results and analysis.} Table \ref{tab:concept-mining} compares our model with different baselines on the UCCM dataset in terms of Exact Match and F1 score.
Results demonstrate that our method achieves the best EM and F1 score.
This is because:
first, the pattern-based concept mining with bootstrapping helps us to construct a collection of high-quality patterns which can accurately extract concepts from queries in an unsupervised manner.
Second, the combination of sequence labeling by CRF and query-title alignment can recognize concepts from both queries and click titles under different situations, i.e., either the concept boundary in query is clear or not.
% In contrast, any isolated strategy will not be able to accurately extract concepts from queries and logs.

We can see the methods based on TextRank \cite{mihalcea2004textrank}, THUCKE \cite{liu2011automatic} and AutoPhrase \cite{shang2018automated} do not give satisfactory performance. That is because existing keyphrases extraction approaches are better suited for extracting keywords or phrases from a long document or a corpus. In contrast, our approach is specially designed for the problem of concept mining from user queries and click titles. Comparing our approach with its variants, including Q-Pattern, Q-CRF, T-CRF and QT-Align, we can see that each component cannot achieve comparable performance as ours independently. This demonstrates the effectiveness of combining different strategies in our system.

\subsection{Evaluation of Document Tagging and Taxonomy Construction}

\textbf{Evaluation of document tagging}. For concept tagging on documents, our system currently processes around 96,700 documents per day, where about 35\% of them can be tagged with at least one concept. We create a dataset containing $11,547$ documents with concept tags for parameter tuning, and we also open-source it for research purpose (see appendix for more details). We evaluate the performance of concept tagging based on this dataset.
The result shows that the precision of concept tagging for documents is $96\%$.
As the correlated concept phrases may even not show in the text, we do not evaluate the recall rate.

\textbf{Evaluation of taxonomy construction}.
We randomly sample 1000 concepts from our taxonomy. As the relationships between \textit{concept-instance} pairs are critical to query and document understanding, our experiment mainly focus on evaluating them. For each concept, we check whether the \textit{isA} relationship between it and its instances is correct. We ask three human judges to evaluate them. For each concept, we record the number of correct instances and the number of incorrect ones.

\begin{table}[tb]
  \caption{Evaluation results of constructed taxonomy.}
  \label{tab:eval-taxonomy}
  \begin{tabular}{ll}
    \toprule
    Metrics / Statistics & Value \\
    \midrule
    Mean \#Instances per Concept & 3.44 \\
    Max \#Instances per Concept & 59 \\
    \textit{isA} Relationship Accuracy & 96.59\% \\
    \bottomrule
  \end{tabular}
  \vspace{-3mm}
\end{table}

Table~\ref{tab:eval-taxonomy} shows the evaluation results. The average number of instances for each concept is $3.44$, and the maximum concept contains $59$ instances. Note that the scale of our taxonomy is keep growing with more daily user queries and query logs.
For the \textit{isA} relationships between \textit{concept-instance} pairs, the accuracy is $96.59\%$.

\begin{table}[tb]
  \caption{Part of the \textit{topic-concept-instance} samples created by \emph{ConcepT} system.}
  \label{tab:show-case}
  \begin{tabularx}{\columnwidth}{>{\hsize=0.22\columnwidth}X|>{\hsize=0.3\columnwidth}X|X}
    \toprule
    Topics & Concepts & Instances \\
    \midrule
    \begin{CJK}{UTF8}{gkai}Entertainment (娱乐)\end{CJK} & \begin{CJK}{UTF8}{gkai}Movies adapted from a novel (小说改编成的电影)\end{CJK} & \begin{CJK}{UTF8}{gkai}The Great Gatsby (了不起的盖茨比)\end{CJK}, \begin{CJK}{UTF8}{gkai}Anna Karenina (安娜·卡列尼娜)\end{CJK}, \begin{CJK}{UTF8}{gkai}Jane Eyre (简爱)\end{CJK} \\
    \begin{CJK}{UTF8}{gkai}Entertainment (娱乐)\end{CJK} & \begin{CJK}{UTF8}{gkai}Female stars with a beautiful smile (笑容最美的女明星)\end{CJK} & \begin{CJK}{UTF8}{gkai}Ayase Haruka (绫濑遥)\end{CJK}, \begin{CJK}{UTF8}{gkai}Sasaki Nozomi (佐佐木希)\end{CJK}, \begin{CJK}{UTF8}{gkai}Dilraba (迪丽热巴)\end{CJK} \\
    \begin{CJK}{UTF8}{gkai}Society \hspace{1cm} (社会)\end{CJK} & \begin{CJK}{UTF8}{gkai}Belt and Road countries along the route (一带一路沿线国家)\end{CJK} & \begin{CJK}{UTF8}{gkai}Palestine (巴勒斯坦)\end{CJK}, \begin{CJK}{UTF8}{gkai}Syria (叙利亚)\end{CJK}, \begin{CJK}{UTF8}{gkai}Mongolia (蒙古)\end{CJK}, \begin{CJK}{UTF8}{gkai}Oman (阿曼)\end{CJK} \\
    \begin{CJK}{UTF8}{gkai}Games   \hspace{1cm}  (游戏)\end{CJK} & \begin{CJK}{UTF8}{gkai}Mobile game for office workers (适合上班族玩的手游)\end{CJK} & \begin{CJK}{UTF8}{gkai}Pokemon (口袋妖怪)\end{CJK}, \begin{CJK}{UTF8}{gkai}Invincible Asia (东方不败)\end{CJK} \\
    \bottomrule
  \end{tabularx}
  \vspace{-5mm}
\end{table}

Table~\ref{tab:show-case} shows a part of \textit{topic-concept-instance} tuples from our taxonomy. We can see that the extracted concepts are expressed from ``user perspective'', such as ``\begin{CJK}{UTF8}{gkai}Female stars with a beautiful smile (笑容最美的女明星)\end{CJK}'' or ``\begin{CJK}{UTF8}{gkai}Mobile game for office workers (适合上班族玩的手游)\end{CJK}''. At the same time, the relationships between concepts and instances are also established based on user activities. For example, when a certain number of users click the documents related to ``\begin{CJK}{UTF8}{gkai}Sasaki Nozomi (佐佐木希)\end{CJK}'' when they are searching ``\begin{CJK}{UTF8}{gkai}Female stars with a beautiful smile (笑容最美的女明星)\end{CJK}'', our system will be able to recognize the \textit{isA} relationship between the concept and the instance.

\subsection{Online A/B Testing for Recommendation}

\begin{figure}[tb]
\centering
\includegraphics[width=3.2in]{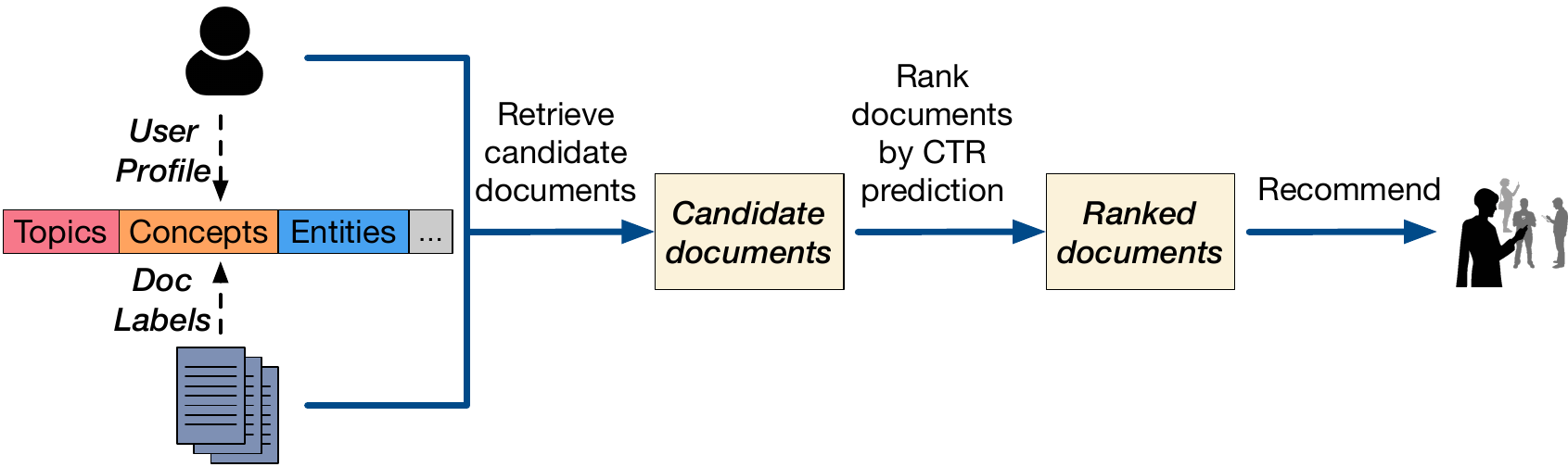}
\vspace{-3mm}
\caption{The framework of feeds recommendation in Tencent QQ Browser.}
\label{fig:Recommend}
\vspace{-3mm}
\end{figure}

We perform large-scale online A/B testing to show how concept tagging on documents helps with improving the performance of recommendation in real world applications.
Fig.~\ref{fig:Recommend} illustrates the recommendation architecture based on our \emph{ConcepT} system. In our system, both users and documents are tagged with interested or related topics, concepts and instances. We first retrieve candidate documents by matching users with documents, then we rank candidate documents by a Click-Through Rate (CTR) prediction model. The ranked documents are pushed to users in the feeds stream of Tencent QQ Browser.

For online A/B testing, we split users into buckets where each bucket contains $800,000$ of users. We first observe and record the activities of each bucket for 7 days based on the following metrics:
\begin{itemize}
  % \item \textbf{Click-Through Rate (CTR)}: the ratio of users who clicked on a specific link to the number of total users who received the link recommendation.
  \item \textbf{Impression Page View (IPV)}: number of pages that matched with users.
  \item \textbf{Impression User View (IUV)}: number of users who has matched pages.
  \item \textbf{Click Page View (CPV)}: number of pages that users clicked.
  \item \textbf{Click User View (CUV)}: number of users who clicked pages.
  \item \textbf{User Conversion Rate (UCR)}: $\frac{CUV}{IUV}$.
  \item \textbf{Average User Consumption (AUC)}: $\frac{CPV}{CUV}$.
  \item \textbf{Users Duration (UD)}: average time users spend on a page.
  \item \textbf{Impression Efficiency (IE)}: $\frac{CPV}{IUV}$.
\end{itemize}
We then select two buckets with highly similar activities. For one bucket, we perform recommendation without the concept tags of documents. For another one, the concept tags of documents are utilized for recommendation. We run our A/B testing for 3 days and compare the result by above metrics. The Impression Efficiency (IE) and Users Duration (UD) are the two most critical metrics in real world application, because they show how many contents users read and how much time they spend on an application.

\begin{table}[tb]
  \caption{Online A/B testing results.}
  \label{tab:ab-testing}
  \begin{tabular}{ll|ll}
    \toprule
    Metrics & Percentage Lift & Metrics & Percentage Lift\\
    \midrule
    IPV & +0.69\% & UCR & +0.04\% \\
    %Click-Through Rate & +0\% \\
    IUV & +0.06\% & AUC & +0.21\% \\
    CPV & +0.38\% & UD & $\mathbf{+0.83\%}$ \\
    CUV & +0.16\% & IE & $\mathbf{+6.01\%}$ \\
    \bottomrule
  \end{tabular}
  \vspace{-3mm}
\end{table}

Table \ref{tab:ab-testing} shows the results of our online A/B testing. In the online experiment, we observe a statistically significant IE gain (6.01\%) and user duration (0.83\%). The page views and user views for click or impression, as well as user conversation rate and average user consumptions, are all improved. These observations prove that the concept tagging for documents greatly benefits the understanding of documents and helps to better match users with their potential interested documents. With the help of user-centered concepts, we can better capture the contained topics in a document even if it does not explicitly mention them. Given more matched documents, users spend more times and reading more articles in our feeds.

\subsection{Offline User Study of Query Rewriting for Searching}

Here we evaluate how user-centered concept mining can help with improving the results of search engines by query rewriting based on conceptualization.
We create a evaluation dataset which contains 108 queries from Tencent QQ Browser.
For each query $\mathbf{q}$, we analyze the concept $\mathbf{c}$ conveyed in the query, and rewrite the query by concatenating each of the instances $\{\mathbf{e}_1, \mathbf{e}_2, \cdots, \mathbf{e}_n\} \in \mathbf{c}$ with $q$. The rewritten queries are in the format of ``$\mathbf{q}$ $\mathbf{e}_i$''.
For the original query, we collect the top 10 search results returned by Baidu search engine, the largest search engine in China.
Assume we replace a query by $K$ different instances. 
We collect top $\lceil\frac{10}{K}\rceil$ search results from Baidu for each of the rewritten queries, combining and keeping 10 of them as the search result after query rewriting.

We ask three human judges to evaluate the relevancy of the results. For each search reuslt, we record majority vote, i.e., ``relevant'' or ``not relevant'', of the human judges, and calculate the percentage of relevance of original queries and rewritten queries.
Our evaluation results show that the percentage of relevant top 10 results increases from $73.1\%$ to $85.1\%$ after rewriting the queries with our strategy.
The reason is that the concept mining for user queries helps to understand the intention of user queries, and concatenating the instances belonging to the concept with the original query provides the search engine more relevant and explicit keywords. Therefore, the search results will better match user's intention.

%!TEX root = main.tex
\section{Related Work}
\label{sec:related}
Our work is mainly related to the following research lines.

\textbf{Concept Mining}.
Existing research work on concept mining mainly relies on predefined linguistic templates, statistical signals, knowledge bases or  concept quality.
Traditional approaches for concept mining are closely related to the work of noun phrase chunking and named entity recognition \cite{nadeau2007survey}.
They either employ heuristics, such as fixed POS-tag patterns, to extract typed entities \cite{ren2017cotype}, or consider the problem as sequence tagging and utilize large-scale labeled training data to train complex deep neural models based on LSTM-CRF or CNN \cite{huang2015bidirectional}.
Another line of work focus on terminology and keyphrase extraction. They extract noun phrases based on  statistical occurrence and co-occurrence signals \cite{frantzi2000automatic}, semantic information from knowledge base \cite{witten2006thesaurus} or  textual features \cite{nguyen2007keyphrase}.
Recent approaches for concept mining rely on phrase quality.
\cite{liu2015mining,shang2018automated} adaptively recognize concepts based on concept quality. They exploit various statistical features such as popularity, informativeness, POS tag sequence ans so on to measure phrase quality, and train the concept quality scoring function by using knowledge base entity names as training labels.
%\cite{liconcept} proposed an embedding based method for generating a set of candidate concepts from corpus, learning their embedding vectors, and evaluating their qualities in the embedding space.

\textbf{Text Conceptualization}.
Conceptualization seeks to map a word or a phrase to a set of concepts as a mechanism of understanding short text such as search queries. Since short text usually lack of context, conceptualization helps better make sense of text data by extending the text with categorical or topical information, and therefore facilitates many applications. \cite{li2007improving} performs query expansion by utilizing Wikipedia as external corpus to understand query for improving ad-hoc retrieval performance. \cite{song2011short} groups instances by their conceptual similarity, and develop a Bayesian inference mechanism to conceptualize each group.  To make further use of context information, \cite{wang2015query} utilize a knowledge base that maps instances to their concepts, and build a knowledge base that maps non-instance words, including verbs and adjectives, to concepts.

\textbf{Relation Extraction}.
Relation Extraction (RE) is to identify  relations between entities and concepts automatically.
% A comprehensive introduction can be found in \cite{pawar2017relation}.
Generally speaking, Relation Extraction techniques can be classified into several categories: 1) supervised techniques including features-based \cite{guodong2005exploring}
and kernel based \cite{culotta2004dependency} methods, 2) semi-supervised approaches including bootstrapping \cite{brin1998extracting}, 3) unsupervised methods \cite{yan2009unsupervised}, 4) Open Information Extraction \cite{fader2011identifying}, and 5) distant supervision based techniques \cite{zeng2015distant}. In our work, we combine unsupervised approaches, semi-supervised bootstrapping technique, and supervised sequence labeling algorithm to extract concepts and identify the relationship between entities and concepts.

%!TEX root = main.tex
\section{Conclusion}
\label{sec:conclude}

In this paper, we describe our experience of implementing \textit{ConcepT}, a user-centered concept mining and tagging system at Tencent that designed to improve the understanding of both queries and long documents.
Our system extracts user-centered concepts from a large amount of user queries and query logs, as well as performs concept tagging on documents to characterize the coverage of documents from user-perspective.
In addition, \emph{ConcepT} further identifies the \textit{isA} relationship between concepts, instances and topics to constructs a 3-layered \textit{topic-concept-instance} taxonomy.
We conduct extensive performance evaluation through both offline experiments and online large-scale A/B testing in the QQ Browser mobile application on more than $800,000$ real users.
The results show that our system can extract featured, user-centered concepts accurately from user queries and query logs, and it is quite helpful for both search engines and recommendation systems. For search engines, the pilot user study in our experiments shows that we improve the results of search engine by query conceptualization. For recommendation, according to the real-world large-scale online A/B testing, the Impression Efficiency improves by 6.01\% when incorporating \emph{ConcepT}system for feeds recommendation in Tencent QQ Browser.

\bibliographystyle{ACM-Reference-Format}
\bibliography{main}

%%% -*-BibTeX-*-
%%% Do NOT edit. File created by BibTeX with style
%%% ACM-Reference-Format-Journals [18-Jan-2012].

\begin{thebibliography}{25}

%%% ====================================================================
%%% NOTE TO THE USER: you can override these defaults by providing
%%% customized versions of any of these macros before the \bibliography
%%% command.  Each of them MUST provide its own final punctuation,
%%% except for \shownote{}, \showDOI{}, and \showURL{}.  The latter two
%%% do not use final punctuation, in order to avoid confusing it with
%%% the Web address.
%%%
%%% To suppress output of a particular field, define its macro to expand
%%% to an empty string, or better, \unskip, like this:
%%%
%%% \newcommand{\showDOI}[1]{\unskip}   % LaTeX syntax
%%%
%%% \def \showDOI #1{\unskip}           % plain TeX syntax
%%%
%%% ====================================================================

\ifx \showCODEN    \undefined \def \showCODEN     #1{\unskip}     \fi
\ifx \showDOI      \undefined \def \showDOI       #1{#1}\fi
\ifx \showISBNx    \undefined \def \showISBNx     #1{\unskip}     \fi
\ifx \showISBNxiii \undefined \def \showISBNxiii  #1{\unskip}     \fi
\ifx \showISSN     \undefined \def \showISSN      #1{\unskip}     \fi
\ifx \showLCCN     \undefined \def \showLCCN      #1{\unskip}     \fi
\ifx \shownote     \undefined \def \shownote      #1{#1}          \fi
\ifx \showarticletitle \undefined \def \showarticletitle #1{#1}   \fi
\ifx \showURL      \undefined \def \showURL       {\relax}        \fi
% The following commands are used for tagged output and should be
% invisible to TeX
\providecommand\bibfield[2]{#2}
\providecommand\bibinfo[2]{#2}
\providecommand\natexlab[1]{#1}
\providecommand\showeprint[2][]{arXiv:#2}

\bibitem[\protect\citeauthoryear{Brin}{Brin}{1998}]%
        {brin1998extracting}
\bibfield{author}{\bibinfo{person}{Sergey Brin}.}
  \bibinfo{year}{1998}\natexlab{}.
\newblock \showarticletitle{Extracting patterns and relations from the world
  wide web}. In \bibinfo{booktitle}{\emph{International Workshop on The World
  Wide Web and Databases}}. Springer, \bibinfo{pages}{172--183}.
\newblock


\bibitem[\protect\citeauthoryear{Culotta and Sorensen}{Culotta and
  Sorensen}{2004}]%
        {culotta2004dependency}
\bibfield{author}{\bibinfo{person}{Aron Culotta} {and} \bibinfo{person}{Jeffrey
  Sorensen}.} \bibinfo{year}{2004}\natexlab{}.
\newblock \showarticletitle{Dependency tree kernels for relation extraction}.
  In \bibinfo{booktitle}{\emph{Proceedings of the 42nd annual meeting on
  association for computational linguistics}}. ACL, \bibinfo{pages}{423}.
\newblock


\bibitem[\protect\citeauthoryear{Fader, Soderland, and Etzioni}{Fader
  et~al\mbox{.}}{2011}]%
        {fader2011identifying}
\bibfield{author}{\bibinfo{person}{Anthony Fader}, \bibinfo{person}{Stephen
  Soderland}, {and} \bibinfo{person}{Oren Etzioni}.}
  \bibinfo{year}{2011}\natexlab{}.
\newblock \showarticletitle{Identifying relations for open information
  extraction}. In \bibinfo{booktitle}{\emph{Proceedings of the conference on
  empirical methods in natural language processing}}. ACL,
  \bibinfo{pages}{1535--1545}.
\newblock


\bibitem[\protect\citeauthoryear{Frantzi, Ananiadou, and Mima}{Frantzi
  et~al\mbox{.}}{2000}]%
        {frantzi2000automatic}
\bibfield{author}{\bibinfo{person}{Katerina Frantzi}, \bibinfo{person}{Sophia
  Ananiadou}, {and} \bibinfo{person}{Hideki Mima}.}
  \bibinfo{year}{2000}\natexlab{}.
\newblock \showarticletitle{Automatic recognition of multi-word terms:. the
  c-value/nc-value method}.
\newblock \bibinfo{journal}{\emph{International journal on digital libraries}}
  \bibinfo{volume}{3}, \bibinfo{number}{2} (\bibinfo{year}{2000}),
  \bibinfo{pages}{115--130}.
\newblock


\bibitem[\protect\citeauthoryear{GuoDong, Jian, Jie, and Min}{GuoDong
  et~al\mbox{.}}{2005}]%
        {guodong2005exploring}
\bibfield{author}{\bibinfo{person}{Zhou GuoDong}, \bibinfo{person}{Su Jian},
  \bibinfo{person}{Zhang Jie}, {and} \bibinfo{person}{Zhang Min}.}
  \bibinfo{year}{2005}\natexlab{}.
\newblock \showarticletitle{Exploring various knowledge in relation
  extraction}. In \bibinfo{booktitle}{\emph{Proceedings of the 43rd annual
  meeting on association for computational linguistics}}. ACL,
  \bibinfo{pages}{427--434}.
\newblock


\bibitem[\protect\citeauthoryear{Hua, Wang, Wang, Zheng, and Zhou}{Hua
  et~al\mbox{.}}{2015}]%
        {hua2015short}
\bibfield{author}{\bibinfo{person}{Wen Hua}, \bibinfo{person}{Zhongyuan Wang},
  \bibinfo{person}{Haixun Wang}, \bibinfo{person}{Kai Zheng}, {and}
  \bibinfo{person}{Xiaofang Zhou}.} \bibinfo{year}{2015}\natexlab{}.
\newblock \showarticletitle{Short text understanding through lexical-semantic
  analysis}. IEEE, \bibinfo{pages}{495--506}.
\newblock


\bibitem[\protect\citeauthoryear{Huang, Xu, and Yu}{Huang
  et~al\mbox{.}}{2015}]%
        {huang2015bidirectional}
\bibfield{author}{\bibinfo{person}{Zhiheng Huang}, \bibinfo{person}{Wei Xu},
  {and} \bibinfo{person}{Kai Yu}.} \bibinfo{year}{2015}\natexlab{}.
\newblock \showarticletitle{Bidirectional LSTM-CRF models for sequence
  tagging}.
\newblock \bibinfo{journal}{\emph{arXiv preprint arXiv:1508.01991}}
  (\bibinfo{year}{2015}).
\newblock


\bibitem[\protect\citeauthoryear{Lehmann, Isele, Jakob, Jentzsch, Kontokostas,
  Mendes, Hellmann, Morsey, Van~Kleef, Auer, et~al\mbox{.}}{Lehmann
  et~al\mbox{.}}{2015}]%
        {lehmann2015dbpedia}
\bibfield{author}{\bibinfo{person}{Jens Lehmann}, \bibinfo{person}{Robert
  Isele}, \bibinfo{person}{Max Jakob}, \bibinfo{person}{Anja Jentzsch},
  \bibinfo{person}{Dimitris Kontokostas}, \bibinfo{person}{Pablo~N Mendes},
  \bibinfo{person}{Sebastian Hellmann}, \bibinfo{person}{Mohamed Morsey},
  \bibinfo{person}{Patrick Van~Kleef}, \bibinfo{person}{S{\"o}ren Auer},
  {et~al\mbox{.}}} \bibinfo{year}{2015}\natexlab{}.
\newblock \showarticletitle{DBpedia--a large-scale, multilingual knowledge base
  extracted from Wikipedia}.
\newblock \bibinfo{journal}{\emph{Semantic Web}} \bibinfo{volume}{6},
  \bibinfo{number}{2} (\bibinfo{year}{2015}), \bibinfo{pages}{167--195}.
\newblock


\bibitem[\protect\citeauthoryear{Li, Luk, Ho, and Chung}{Li
  et~al\mbox{.}}{2007}]%
        {li2007improving}
\bibfield{author}{\bibinfo{person}{Yinghao Li}, \bibinfo{person}{Wing
  Pong~Robert Luk}, \bibinfo{person}{Kei Shiu~Edward Ho}, {and}
  \bibinfo{person}{Fu~Lai~Korris Chung}.} \bibinfo{year}{2007}\natexlab{}.
\newblock \showarticletitle{Improving weak ad-hoc queries using wikipedia
  asexternal corpus}. In \bibinfo{booktitle}{\emph{SIGIR}}. ACM,
  \bibinfo{pages}{797--798}.
\newblock


\bibitem[\protect\citeauthoryear{Liu, Shang, Wang, Ren, and Han}{Liu
  et~al\mbox{.}}{2015}]%
        {liu2015mining}
\bibfield{author}{\bibinfo{person}{Jialu Liu}, \bibinfo{person}{Jingbo Shang},
  \bibinfo{person}{Chi Wang}, \bibinfo{person}{Xiang Ren}, {and}
  \bibinfo{person}{Jiawei Han}.} \bibinfo{year}{2015}\natexlab{}.
\newblock \showarticletitle{Mining quality phrases from massive text corpora}.
  In \bibinfo{booktitle}{\emph{SIGMOD}}. ACM, \bibinfo{pages}{1729--1744}.
\newblock


\bibitem[\protect\citeauthoryear{Liu, Chen, Zheng, and Sun}{Liu
  et~al\mbox{.}}{2011}]%
        {liu2011automatic}
\bibfield{author}{\bibinfo{person}{Zhiyuan Liu}, \bibinfo{person}{Xinxiong
  Chen}, \bibinfo{person}{Yabin Zheng}, {and} \bibinfo{person}{Maosong Sun}.}
  \bibinfo{year}{2011}\natexlab{}.
\newblock \showarticletitle{Automatic keyphrase extraction by bridging
  vocabulary gap}. In \bibinfo{booktitle}{\emph{Proceedings of the Fifteenth
  Conference on Computational Natural Language Learning}}. ACL,
  \bibinfo{pages}{135--144}.
\newblock


\bibitem[\protect\citeauthoryear{Mihalcea and Tarau}{Mihalcea and
  Tarau}{2004}]%
        {mihalcea2004textrank}
\bibfield{author}{\bibinfo{person}{Rada Mihalcea} {and} \bibinfo{person}{Paul
  Tarau}.} \bibinfo{year}{2004}\natexlab{}.
\newblock \showarticletitle{Textrank: Bringing order into text}. In
  \bibinfo{booktitle}{\emph{EMNLP}}.
\newblock


\bibitem[\protect\citeauthoryear{Nadeau and Sekine}{Nadeau and Sekine}{2007}]%
        {nadeau2007survey}
\bibfield{author}{\bibinfo{person}{David Nadeau} {and} \bibinfo{person}{Satoshi
  Sekine}.} \bibinfo{year}{2007}\natexlab{}.
\newblock \showarticletitle{A survey of named entity recognition and
  classification}.
\newblock \bibinfo{journal}{\emph{Lingvisticae Investigationes}}
  \bibinfo{volume}{30}, \bibinfo{number}{1} (\bibinfo{year}{2007}),
  \bibinfo{pages}{3--26}.
\newblock


\bibitem[\protect\citeauthoryear{Nguyen and Kan}{Nguyen and Kan}{2007}]%
        {nguyen2007keyphrase}
\bibfield{author}{\bibinfo{person}{Thuy~Dung Nguyen} {and}
  \bibinfo{person}{Min-Yen Kan}.} \bibinfo{year}{2007}\natexlab{}.
\newblock \showarticletitle{Keyphrase extraction in scientific publications}.
  In \bibinfo{booktitle}{\emph{ICADL}}. Springer, \bibinfo{pages}{317--326}.
\newblock


\bibitem[\protect\citeauthoryear{Ren, Wu, He, Qu, Voss, Ji, Abdelzaher, and
  Han}{Ren et~al\mbox{.}}{2017}]%
        {ren2017cotype}
\bibfield{author}{\bibinfo{person}{Xiang Ren}, \bibinfo{person}{Zeqiu Wu},
  \bibinfo{person}{Wenqi He}, \bibinfo{person}{Meng Qu},
  \bibinfo{person}{Clare~R Voss}, \bibinfo{person}{Heng Ji},
  \bibinfo{person}{Tarek~F Abdelzaher}, {and} \bibinfo{person}{Jiawei Han}.}
  \bibinfo{year}{2017}\natexlab{}.
\newblock \showarticletitle{Cotype: Joint extraction of typed entities and
  relations with knowledge bases}. In \bibinfo{booktitle}{\emph{WWW}}.
  International World Wide Web Conferences Steering Committee,
  \bibinfo{pages}{1015--1024}.
\newblock


\bibitem[\protect\citeauthoryear{Shang, Liu, Jiang, Ren, Voss, and Han}{Shang
  et~al\mbox{.}}{2018}]%
        {shang2018automated}
\bibfield{author}{\bibinfo{person}{Jingbo Shang}, \bibinfo{person}{Jialu Liu},
  \bibinfo{person}{Meng Jiang}, \bibinfo{person}{Xiang Ren},
  \bibinfo{person}{Clare~R Voss}, {and} \bibinfo{person}{Jiawei Han}.}
  \bibinfo{year}{2018}\natexlab{}.
\newblock \showarticletitle{Automated phrase mining from massive text corpora}.
\newblock \bibinfo{journal}{\emph{IEEE Transactions on Knowledge and Data
  Engineering}} \bibinfo{volume}{30}, \bibinfo{number}{10}
  (\bibinfo{year}{2018}), \bibinfo{pages}{1825--1837}.
\newblock


\bibitem[\protect\citeauthoryear{Song, Shi, Li, and Zhang}{Song
  et~al\mbox{.}}{2018}]%
        {N18-2028}
\bibfield{author}{\bibinfo{person}{Yan Song}, \bibinfo{person}{Shuming Shi},
  \bibinfo{person}{Jing Li}, {and} \bibinfo{person}{Haisong Zhang}.}
  \bibinfo{year}{2018}\natexlab{}.
\newblock \showarticletitle{Directional Skip-Gram: Explicitly Distinguishing
  Left and Right Context for Word Embeddings}. In
  \bibinfo{booktitle}{\emph{Proceedings of the 2018 Conference of the North
  American Chapter of the Association for Computational Linguistics: Human
  Language Technologies, Volume 2 (Short Papers)}}. \bibinfo{publisher}{ACL},
  \bibinfo{pages}{175--180}.
\newblock
\urldef\tempurl%
\url{https://doi.org/10.18653/v1/N18-2028}
\showDOI{\tempurl}


\bibitem[\protect\citeauthoryear{Song, Wang, Wang, Li, and Chen}{Song
  et~al\mbox{.}}{2011}]%
        {song2011short}
\bibfield{author}{\bibinfo{person}{Yangqiu Song}, \bibinfo{person}{Haixun
  Wang}, \bibinfo{person}{Zhongyuan Wang}, \bibinfo{person}{Hongsong Li}, {and}
  \bibinfo{person}{Weizhu Chen}.} \bibinfo{year}{2011}\natexlab{}.
\newblock \showarticletitle{Short text conceptualization using a probabilistic
  knowledgebase}. In \bibinfo{booktitle}{\emph{Proceedings of the twenty-second
  international joint conference on artificial intelligence-volume volume
  three}}. AAAI Press, \bibinfo{pages}{2330--2336}.
\newblock


\bibitem[\protect\citeauthoryear{Suchanek, Kasneci, and Weikum}{Suchanek
  et~al\mbox{.}}{2007}]%
        {suchanek2007yago}
\bibfield{author}{\bibinfo{person}{Fabian~M Suchanek}, \bibinfo{person}{Gjergji
  Kasneci}, {and} \bibinfo{person}{Gerhard Weikum}.}
  \bibinfo{year}{2007}\natexlab{}.
\newblock \showarticletitle{Yago: a core of semantic knowledge}. In
  \bibinfo{booktitle}{\emph{Proceedings of the 16th international conference on
  World Wide Web}}. ACM, \bibinfo{pages}{697--706}.
\newblock


\bibitem[\protect\citeauthoryear{Wang, Zhao, Wang, Meng, and Wen}{Wang
  et~al\mbox{.}}{2015}]%
        {wang2015query}
\bibfield{author}{\bibinfo{person}{Zhongyuan Wang}, \bibinfo{person}{Kejun
  Zhao}, \bibinfo{person}{Haixun Wang}, \bibinfo{person}{Xiaofeng Meng}, {and}
  \bibinfo{person}{Ji-Rong Wen}.} \bibinfo{year}{2015}\natexlab{}.
\newblock \showarticletitle{Query Understanding through Knowledge-Based
  Conceptualization}. In \bibinfo{booktitle}{\emph{IJCAI}}.
  \bibinfo{pages}{3264--3270}.
\newblock


\bibitem[\protect\citeauthoryear{Witten and Medelyan}{Witten and
  Medelyan}{2006}]%
        {witten2006thesaurus}
\bibfield{author}{\bibinfo{person}{Ian~H Witten} {and} \bibinfo{person}{Olena
  Medelyan}.} \bibinfo{year}{2006}\natexlab{}.
\newblock \showarticletitle{Thesaurus based automatic keyphrase indexing}. In
  \bibinfo{booktitle}{\emph{Proceedings of the 6th ACM/IEEE-CS Joint Conference
  on Digital Libraries (JCDL'06)}}. IEEE, \bibinfo{pages}{296--297}.
\newblock


\bibitem[\protect\citeauthoryear{Wu, Li, Wang, and Zhu}{Wu
  et~al\mbox{.}}{2012}]%
        {wu2012probase}
\bibfield{author}{\bibinfo{person}{Wentao Wu}, \bibinfo{person}{Hongsong Li},
  \bibinfo{person}{Haixun Wang}, {and} \bibinfo{person}{Kenny~Q Zhu}.}
  \bibinfo{year}{2012}\natexlab{}.
\newblock \showarticletitle{Probase: A probabilistic taxonomy for text
  understanding}. In \bibinfo{booktitle}{\emph{Proceedings of the 2012 ACM
  SIGMOD International Conference on Management of Data}}. ACM,
  \bibinfo{pages}{481--492}.
\newblock


\bibitem[\protect\citeauthoryear{Yan, Okazaki, Matsuo, Yang, and Ishizuka}{Yan
  et~al\mbox{.}}{2009}]%
        {yan2009unsupervised}
\bibfield{author}{\bibinfo{person}{Yulan Yan}, \bibinfo{person}{Naoaki
  Okazaki}, \bibinfo{person}{Yutaka Matsuo}, \bibinfo{person}{Zhenglu Yang},
  {and} \bibinfo{person}{Mitsuru Ishizuka}.} \bibinfo{year}{2009}\natexlab{}.
\newblock \showarticletitle{Unsupervised relation extraction by mining
  wikipedia texts using information from the web}. In
  \bibinfo{booktitle}{\emph{Proceedings of the Joint Conference of the 47th
  Annual Meeting of the ACL and the 4th International Joint Conference on
  Natural Language Processing of the AFNLP: Volume 2-Volume 2}}. ACL,
  \bibinfo{pages}{1021--1029}.
\newblock


\bibitem[\protect\citeauthoryear{Zeng, Liu, Chen, and Zhao}{Zeng
  et~al\mbox{.}}{2015}]%
        {zeng2015distant}
\bibfield{author}{\bibinfo{person}{Daojian Zeng}, \bibinfo{person}{Kang Liu},
  \bibinfo{person}{Yubo Chen}, {and} \bibinfo{person}{Jun Zhao}.}
  \bibinfo{year}{2015}\natexlab{}.
\newblock \showarticletitle{Distant supervision for relation extraction via
  piecewise convolutional neural networks}. In
  \bibinfo{booktitle}{\emph{EMNLP}}. \bibinfo{pages}{1753--1762}.
\newblock


\bibitem[\protect\citeauthoryear{Zheng, Chen, Sun, and Zha}{Zheng
  et~al\mbox{.}}{2007}]%
        {zheng2007regression}
\bibfield{author}{\bibinfo{person}{Zhaohui Zheng}, \bibinfo{person}{Keke Chen},
  \bibinfo{person}{Gordon Sun}, {and} \bibinfo{person}{Hongyuan Zha}.}
  \bibinfo{year}{2007}\natexlab{}.
\newblock \showarticletitle{A regression framework for learning ranking
  functions using relative relevance judgments}. In
  \bibinfo{booktitle}{\emph{SIGIR}}. ACM, \bibinfo{pages}{287--294}.
\newblock


\end{thebibliography}

\clearpage
%!TEX root = main.tex
\appendix

\section{Information for Reproducibility}

\subsection{System Implementation and Deployment}

We implement and deploy our \emph{ConcepT} system in Tencent QQ Browser.
The concept mining module and taxonomy construction module are implemented in Python 2.7, and they run as offline components.
For document tagging module, it is implemented in C++ and runs as an online service.
We utilize MySQL for data storage.

In our system, each component works as a service and is deployed on Tencent Total Application Framework (Tencent TAF)\footnote{https://github.com/TarsCloud/Tars}. Tencent TAF is a high-performance remote procedure call (RPC) framework based on name service and Tars protocol, it also integrate administration platform, and implement hosting-service via flexible schedule. It has been used in Tencent since 2008, and supports different programming languages.
For online document concept tagging, it is running on 50 dockers. Each docker is configured with six 2.5 GHz Intel Xeon Gold 6133 CPU cores and 6 GB memory.
For offline concept mining and taxonomy construction, they are running on 2 dockers with the same configuration.

\begin{algorithm}[tbh]
 \KwData{Queries and query logs in a day}
 \KwResult{Concepts}
 Check whether successfully obtained the queries and logs\;
 \eIf{succeed}{
   Perform concept mining by our proposed approach\;
   }{
   Break\;
 }
 \caption{Offline concept mining process.}
\end{algorithm}

\begin{algorithm}[tbh]
 \KwData{News documents, the vocabulary of instances, concepts, and the index between key terms and concepts}
 \KwResult{\textit{isA} relationship between concepts and instances}
 \For{each document}{
	 Get the instances in the document based on the vocabulary\;
	 \For{each instance}{
	 Get the intersection of concept key terms and the terms co-occurred in the same sentence with document instances\;
	 Get related concepts that containing at least one key term in the intersection\;
	 Get <instance, key terms, concepts> tuples based on the results of above steps\;
	 }
 }
 Get the co-occurrence features listed in Table \ref{tab:features}, and classify whether existing \textit{isA} relationship between the instances and candidate concepts.

 \caption{Offline \textit{isA} relationship discovery between concepts and instances.}
\end{algorithm}

\begin{algorithm}[tbh]
 \KwData{News documents, \textit{isA} relationship between instances and concepts}
 \KwResult{Documents with concept tags}
 \For{each document}{
	 Perform word segmentation\;
	 Extract key instances by the approach described in Fig.~\ref{fig:DocConcept}\;
     Get candidate concepts by the \textit{isA} relationship between concepts and key instances\;
	 \For{each concept}{
	 Calculate the coherence between the candidate concept and the document by the probabilistic inference-based approach\;
	 Tag the concept to the document if the coherence is above a threshold\;
	 }
 }

 \caption{Online probabilistic inference-based concept tagging for documents.}
\end{algorithm}

\begin{algorithm}[tbh]
 \KwData{News documents}
 \KwResult{Documents with concept tags}
 \For{each document}{
	 Perform word segmentation\;
	 Extract key terms by TF-IDF\;
     Get candidate concepts containing above key terms\;
     Get the title-enriched representation of candidate concepts\;
     Represent document and each candidate concept by TF-IDF vector\;
	 \For{each concept}{
	 Calculate cosine similarity between the candidate concept and the document\;
	 Tag the concept to the document if the similarity is above a threshold\;
	 }
 }

 \caption{Online matching-based concept tagging for documents.}
\end{algorithm}

Algorithm 1-4 show the running processes of each component in ConcepT.
For offline concept mining from queries and search logs, the component is running on a daily basis. It extracts around 27,000 concepts from 25 millions of query logs everyday, and about 11,000 of the extracted concepts are new.
For offline relationship discovery in taxonomy construction, the component runs every two weeks.
For online concept tagging for documents, the processing speed is 40 documents per second. It performs concept tagging for about 96,700 documents per day, where about 35\% of them can be tagged with at least one concept.

\subsection{Parameter Settings and Training Process}

We have described the threshold parameters in our paper.
Here we introduce the features we use for different components in our system, and describe how we train each component.
Table \ref{tab:features} lists the input features we use for different sub-modules in our \emph{ConcepT} system.

\begin{table}[tb]
  \caption{The features we use for different tasks in ConcepT.}
  \label{tab:features}
  \begin{tabularx}{\columnwidth}{>{\hsize=0.3\hsize}X|X}
    \toprule
    Task & Features\\
    \midrule
    Document key instance extraction & Whether the topic of instance is the same with the topic of document; whether it is the same with the instance in title; whether the title contains the instance topic; the frequency of the instance among all instances in the document; the percentage of sentences containing the instance.\\
    \hline
    Classify whether a short text is a concept & Whether the short text ever shown as a user query; how many times it has been searched; Bag-of-Word representation of the text; the topic distribution of user clicked documents given that short text as query.\\
    \hline
    Train CRF for concept mining from query & word, NER, POS, <previous word, word>, <previous word, next word>, <previous POS, POS>, <POS, next POS>, <previous POS, word>, <word, next POS>. \\
    \bottomrule
  \end{tabularx}
  \vspace{0mm}
\end{table}

\textbf{Training process.} 
For concept mining, we randomly sample 15,000 query search logs in Tencent QQ Browser within one month.
We extract concepts for these query logs using approaches introduced in Sec. \ref{sec:modelquery}, and the results are manually checked by Tencent product managers.
The resulting dataset is used to train the classifier in query-title alignment-based concept mining, and the Conditional Random Field in our model. We utilize CRF++ v0.58 to train our model. 80\% of the dataset is used as training set, 10\% as development set and the remaining 10\% as test set.

For concept tagging, we randomly sample 10,000 news articles from the feeds stream of Tencent QQ browser during a three-month period, where each topic contains abut 800 to 1000 articles.
We iteratively perform concept tagging for documents based on the approaches described in Sec. \ref{sec:modeldoc}. After each iteration, we manually check whether the tagged concepts are correlated or not. Then we update our dataset and retrain the models of concept tagging. The iteration process is topped until no more new concepts can be tagged to documents.
The resulting dataset is used to train the classifiers and set the hyper-parameters in concept tagging. We use 80\% of the dataset as training set, 10\% as development set and the remaining 10\% as test set.

\subsection{Publish Our Datasets}

We have published our datasets for research purpose and they can be accessed from \url{https://github.com/BangLiu/ConcepT}. Specifically, we open source the following datasets:
\begin{itemize}
  \item \textbf{The UCCM dataset}. It is used to evaluate the performance of our approach for concept mining and it contains $10,000$ samples.
  \item \textbf{The document tagging dataset}. It is used to evaluate the document tagging accuracy of ConcepT, and it contains 11,547 documents with concept tags.
  \item \textbf{Topic-concept-instance taxonomy}. It contains 1000 \textit{topic-concept-instance} samples from our constructed taxonomy.
  \item \textbf{The seed concept patterns for bootstrapping-based concept mining}. It contains the seed string patterns we utilized for bootstrapping-based concept mining from queries.
  \item \textbf{Pre-defined topic list}. It contains our $31$ pre-defined topics for taxonomy construction.
\end{itemize}

\subsection{Details about Document Topic Classification}

% \begin{figure}[tb]
% \centering
% \includegraphics[width=3.3in]{KeywordExtraction}
% \vspace{-3mm}
% \caption{Extracting keywords from documents.}
% \label{fig:KeywordExtraction}
% \vspace{-3mm}
% \end{figure}

\begin{figure}[tb]
\centering
\includegraphics[width=2.4in]{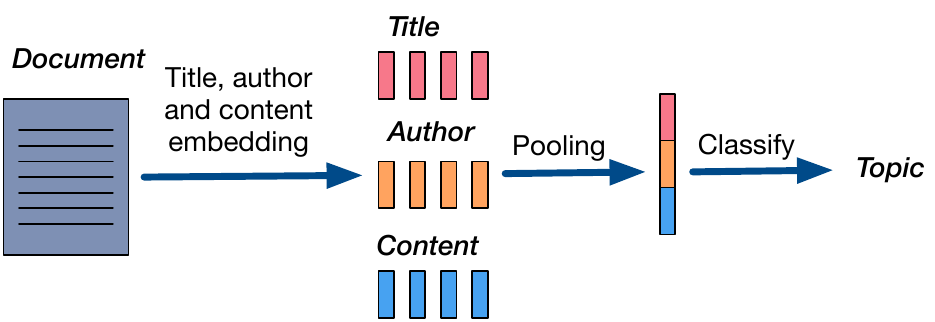}
\vspace{-3mm}
\caption{Document topic classification.}
\label{fig:TopicClassification}
\vspace{-3mm}
\end{figure}

Topic classification aims to classify a document $d$ into our predefined 
$N_t$ (it is 31 in our system) topic categories, including entertainment, events, technology and so forth.
Fig.~\ref{fig:TopicClassification} illustrates our model for document topic classification. We represent the title, author, and content of document $d$ by word vectors.
Then we apply max pooling to title and author embeddings, and mean pooling to content embeddings. The results of pooling operations are concatenated into a fix-length vector representation of $d$. We then classify it by a feed forward neural network. The accuracy of our model is 95\% on a labeled dataset containing 35,000 news articles.

\subsection{Examples of Queries and Extracted Concepts}

\begin{table}[tb]
  \caption{Examples of queries and the extracted concepts given by ConcepT.}
  \label{tab:query-showcase}
  \begin{tabularx}{\columnwidth}{X|X}
    \toprule
    Query & Concept\\
    \midrule
    What are the Qianjiang specialties \begin{CJK}{UTF8}{gkai}(黔江的特产有哪些)\end{CJK} & Qianjiang specialties \begin{CJK}{UTF8}{gkai}(黔江特产)\end{CJK} \\
    Collection of noodle snacks cooking methods \begin{CJK}{UTF8}{gkai}(面条小吃的做法大全)\end{CJK} & noodle snacks cooking methods \begin{CJK}{UTF8}{gkai}(面条小吃的做法)\end{CJK} \\
    Which cars are cheap and fuel-efficient? \begin{CJK}{UTF8}{gkai}(有什么便宜省油的车)\end{CJK} & cheap and fuel-efficient cars \begin{CJK}{UTF8}{gkai}(便宜省油的车)\end{CJK} \\
    Jiyuan famous snacks \begin{CJK}{UTF8}{gkai}(济源有名的小吃)\end{CJK} & Jiyuan snacks \begin{CJK}{UTF8}{gkai}(济源小吃)\end{CJK} \\
    What are the symptoms of depression? \begin{CJK}{UTF8}{gkai}(抑郁症有什么症状)\end{CJK} & symptoms of depression \begin{CJK}{UTF8}{gkai} (抑郁症症状)\end{CJK} \\
    Large-scale games of military theme \begin{CJK}{UTF8}{gkai}(军事题材的大型游戏)\end{CJK} & Military games \begin{CJK}{UTF8}{gkai}(军事游戏)\end{CJK} \\
    \bottomrule
  \end{tabularx}
  \vspace{0mm}
\end{table}

% \begin{figure}[!ht]
% \centering
% \includegraphics[width=0.4\textwidth]{showcase}
% %\vspace{-5mm}
% \caption{Show cases of concept tagging for documents.}
% \label{fig:showcase}
% %\vspace{-5mm}
% \end{figure}

% Here we show some results of different components, so that readers can know what can be obtained with our system.

Table \ref{tab:query-showcase} lists some examples of user queries, together with the concepts extracted by ConcepT. We can see that the concepts are appropriate to summarize the core user intention in queries.

% Fig.~\ref{fig:showcase} shows four examples in the feeds of Tencent QQ Broswer. Our system is able to tag highly coherent concepts with documents of different topics, such as housing, entertainment, food, game and so on.

\end{document}